\documentclass[10pt,a4paper,twocolumn,prd]{revtex4}
\usepackage{amsfonts}
\usepackage{amsmath}
\usepackage{amssymb}
\usepackage{graphicx}
\usepackage{fancyhdr}

\input{tcilatex}
\begin{document}

\title{\textbf{Particle decay processes, the quantum Zeno effect and the
continuity of time}}
\author{George Jaroszkiewicz and Jon Eakins}

\begin{abstract}
Signal-state quantum mechanics is used to discuss quantum mechanical
particle decay probabilities and the quantum Zeno effect. This approach
avoids the assumption of continuous time, conserves total probability and
requires neither non-Hermitian Hamiltonians nor the ad-hoc introduction of
complex energies. The formalism is applied to single channel decays, the
ammonium molecule, and neutral Kaon decay processes.

\

\noindent \textbf{KEYWORDS:} {\ particle decays, quantum Zeno effect,
quantum bits, neutral Kaon decay, temporal continuity}

\

\noindent \textbf{PACS:} 03.65.Ca, 03.65.Ta
\end{abstract}

\maketitle

\address{School of Mathematical Sciences, University of Nottingham, \\
University Park, Nottingham NG72RD, UK}


\section{Introduction}

The nature of time is not generally regarded as one of the fundamental
problems in quantum mechanics (QM). It is normally assumed to be a
continuum. Temporal continuity is central to the formalism of classical
mechanics (CM) and relativity, and many of the concepts derived from it have
analogues in quantum theory. For example, the Hamiltonian is a generator of
translation in time in both CM and QM. In standard quantum mechanics (SQM),
the empirical success of the Schr\"{o}dinger equation is usually taken as
supporting temporal continuity, although there have been many attempts to
construct discrete time quantum mechanics (Caldirola 1978, Yamamoto 1984,
Bender et al. 1985, Golden 1992, Jaroszkiewicz and Norton 1997).
Significantly, all successful quantum field theories such as QED and the
standard model are based on continuous time. If temporal continuity were
removed, the ramifications would be enormous in a number of disciplines.

On closer inspection, however, the continuity of time does not look quite so
obvious. A problem emerges, involving a clash between two contrasting and
essentially mutually exclusive observations. These were discussed in detail
by Misra and Sudarshan (M\&S) in an influential paper on the quantum Zeno
effect (Misra and Sudarshan 1977). On the one hand, there is currently no
known principle in SQM which forbids the continuity of time. Because of
this, the axioms of SQM are invariably stated in terms of continuous time
and usually include the Schr\"{o}dinger equation explicitly (Peres 1993). On
the other hand, there is the empirical fact that no experiment can ever
monitor a system in a continuous way. The best that can be done is to
perform a sequence of experiments with a decreasing measurement time scale,
in an attempt to see evidence for temporal continuity, such as in the
phenomenon known as the quantum Zeno effect (Bollinger et al. 1990).

The standard response to this problem is to assert that, because experiments
\emph{could in principle} be conducted with an arbitrarily small measurement
time scale $\tau $, there is in fact no barrier to taking the limit $\tau
\rightarrow 0$. This argument is unsound because all the theoretical and
empirical evidence actually points in the opposite direction. On the
theoretical side, Caldirola (1978) proposed a model for the electron based
on the existence of a fundamental time scale, the chronon, of the order $%
10^{-23}$ seconds. On a different front, data from high energy particle
scattering experiments indicates that at a certain scale of energy, the
weak, electromagnetic and strong interaction coupling constants run to a
common limit, \noindent a phenomenon known as Grand Unification. The scale
of energy involved corresponds to a timescale of about $10^{-33}$ seconds.
When gravity is included in an attempt to unify all known interactions, the
relevant timescale reduces to the Planck time, of the order $10^{-44}$
seconds. This is popularly regarded as representing the smallest meaningful
temporal interval possible, although that belief itself has been questioned
(Meschini 2006).

The Grand Unification and Planck timescales are generally regarded as most
relevant to early universe cosmology, although there are ongoing attempts in
high energy particle physics scattering experiments to push interaction
energies towards the Grand Unification scale. These timescales however are
far removed from any direct measurement timescales relevant to laboratory
physics. The smallest unit of time that has been measured in the laboratory
is on the attosecond $(10^{-18})$ timescale. Even if this were improved on,
long before Grand Unification or Planck scales could be reached, the
nucleonic timescale of the order $10^{-23}$ seconds (about the time for
light to cross a proton and coincidentally about the same as Caldirola's
chronon) would step in to provide a realistic temporal barrier to the
continuum limit. This is almost certainly the shortest timescale that could
be achieved in any conceivable experiment on the quantum Zeno effect. What
characterizes quantum Zeno effect experiments is the need to probe a system
repeatedly, in an attempt to simulate a continuum of observations. That is
quite different to what happens in a high energy particle collision
experiment, which can be regarded as a succession of one-off observations
per run.

The conventional belief in temporal continuity, therefore, is no more than
an article of faith, a temporal equivalent of Fourier's principle of
similitude (Wigner, 1949), which is known to be false. This principle
asserts that the physical properties of a system is independent of its
scale, and is flatly contradicted by the existence of atoms and the finite
speed of light. Temporal continuity is a metaphysical (i.e., unprovable)
assumption best avoided if at all possible, which is one of the principles
on which our work is based.

M\&S discussed a number of themes related to the nature of time. They
analyzed particle decay processes and certain questions not normally
discussed in SQM. Three of these questions were referred to as $P$, $Q$ and $%
R$ and this convention will be followed here. The question $P(0,t;\rho )$
asks for the probability that an unstable system prepared at time zero in
state $\rho $ has decayed sometime during the interval $[0,t]$; the question
$Q(0,t;\rho )$ asks for the probability that the prepared state has not
decayed during this interval; and finally, the question $R(0,t_{1},t;\rho )$
asks for the probability that the state has not decayed during the interval $%
[0,t_{1}]$, where $0<t_{1}<t$, and has decayed during the interval $%
[t_{1},t] $. M\&S stressed that these questions are not what SQM normally
calculates, which is the probability distribution of the time at which decay
occurs, denoted by $q$. The difference here is that the $P$,$\ Q$ and $R$
questions involve a continuous set of observations, or the nearest practical
equivalent of it, whereas $q$ involves a set of repeated runs, each with a
one-off observation at a different time to determine whether the particle
has decayed or not at that time. Because $P$,$\ Q$ and $R$ involve a
different experimental protocol to $q$, it should be expected that empirical
differences will be observed. Note that the observations M\&S refer to can
have negative outcomes, i.e., an answer that a particle has \emph{not}
decayed by a certain time counts as an observation.

The points raised by M\&S were used by them to emphasize the limitations of
SQM. Although it works excellently in all known situations, SQM does not
readily give a complete picture of experimental questions such as $P$, $Q$
and $R$. M\&S discussed a range of alternative resolutions to the questions
raised, concluding that $``$there is no standard and detailed theory for the
actual coupling between quantum systems and the \emph{classical measuring
apparatus}$\textquotedblright $.

Our paper presents an approach to quantum mechanics, \emph{signal-state
quantum mechanics} \footnote{%
In earlier papers, our formalism was referred to as $``$system-free$%
\textquotedblright $ quantum mechanics, but this has been changed here to
avoid any undue association with metaphysics.}(SSQM), which addresses some
of the issues raised by M\&S (Jaroszkiewicz and Ridgway-Taylor 2006,
Jaroszkiewicz and Eakins 2006). In SSQM, quantum wave-functions are
interpreted as probability amplitudes for signals obtained from physical
apparatus by observers. This is in contrast to SQM, in which quantum
wave-functions are generally assumed to describe the properties of systems
under observation, such as electrons, photons, atoms or molecules.

SSQM was motivated greatly by Heisenberg's original vision of quantum
mechanics, in which only quantities accessible to an observer are regarded
as physically meaningful (Heisenberg 1927). Its mathematical structure has
been designed to reflect the fact that experimentalists never deal directly
with particles per se. Instead, they push buttons, look at screens and count
signals, and by such means extract information from their apparatus.
Everything else is inferred.

In SSQM, some concepts from quantum information theory are used to model
states of an observer's apparatus, which is assumed to consist of a varying
number of elementary detectors. Each of these has the potential to provide
the observer with a yes/no answer to the basic question $``$\emph{is there a
signal here at this time?}$\textquotedblright $\emph{\ }M\&S touched upon
this aspect by working with a projection operator $E$ applied at a given
time to an evolving unstable particle state, corresponding to the yes/no
question $``$\emph{is the particle in its undecayed state at this time?}$%
\textquotedblright $. Crucially, M\&S included the decay product states in
their Hilbert space, and this also has an analogue in the SSQM approach.

An important feature in SSQM is that the observer need not choose to look at
any given detector at any given time, and it is here that the quantum nature
of an experiment manifests itself. At any given time, the observer's
information about the potential state of the apparatus can be a
non-classical superposition of alternative classical signal states, prior to
any question being actually asked. Such a state will be called a \emph{%
labstate}, to distinguish it from the concept of system state as used in SQM.

When one or more elementary questions are actually asked and answered, this
inevitably requires the observer to physically interact with the
corresponding detectors. The result is the extraction of physical
information in the form of yes or no answers, and this necessarily modifies
the observer's knowledge of the labstate. It is the labstate which can be
said to collapse, rather than any wave-function of the system under
observation. By placing emphasis on the apparatus and avoiding any focus on
system states, SSQM avoids metaphysical speculation about the relationship
between particles, waves and non-classical particle trajectories.

It is a basic premise in SSQM that all physics experiments can be formulated
in terms of sufficiently many elementary questions and answers. All the
standard concepts of SQM are to be found in SSQM but in a modified and
usually enhanced form. There are some important additional concepts which
arise because of the need to model laboratory physics more realistically
than is the rule in SQM. Particle decay experiments provide an excellent
forum to discuss these ideas, because such experiments are generally much
more complicated in their space-time structure than the simple SQM
description implies.

One of the consequences of this simplification is that when decay products
are not included in the Hilbert space in SQM, it usually becomes necessary
to introduce imaginary terms into effective Hamiltonians, in order to
reflect the inevitable non-conservation of probability which arises in
consequence. Crucially, M\&S did not make this simplification and so their
formalism did not invoke any such assumptions. However, this left them with
significant mathematical problems, which they solved only by making
conventional assumptions concerning evolution operators and the continuity
of time. For example, they assumed that the standard evolution operator $%
U(t)\equiv \exp \left( -iHt/\hbar \right) $ is an element of a strongly
continuous one-parameter family of unitary operators in a separable Hilbert
space. In the SSQM formalism, there is a different Hilbert space associated
with each time step, so such assumptions do not have any natural place.

In the SSQM approach, total probability is always conserved. This is
achieved through the use of an analogue of unitarity called \emph{%
semi-unitarity}. This is not just a technical trick but an essential feature
of the theory. It will be shown that semi-unitarity ensures that overall
probability is conserved and that exponentially falling survival
probabilities can arise in a natural way.

In SSQM, the number of qubits required to describe an experiment can change
with time. At any given time, the qubits needed to describe the apparatus
form what is called a \emph{Heisenberg net}. This is discussed in detail in
the next section. In particle decay processes, it is found that the size of
the Heisenberg net required to describe such processes grows linearly with
time, reflecting the basic fact that a particle decay experiment is an
irreversible quantum process occurring over extended regions of space-time.
In such processes, information can be extracted not only at the end of a
run, but at any time between the start (state preparation) and the end.

The mathematical basis of SSQM will be reviewed briefly in the next section,
followed by a discussion of how it can be used to describe the simplest
idealized decay process, that of a particle decaying via one channel only.
The quantum Zeno effect makes an appearance at this point, and it will be
shown that there is no unambiguous answer as to whether a system decays
whilst it is being monitored or whether it remains in its initial state.

One of the aims of this paper is to show how more complex phenomena such as
neutral Kaon decay can be discussed in SSQM. To do this, it will be
necessary to review the SSQM description of the ammonium molecule. The
techniques developed there are then used to show how the Kaon decay
regeneration amplitudes discussed by Gell-Mann and Pais arise naturally in
the SSQM approach. It will be seen not to be necessary to introduce any
ad-hoc imaginary terms in any amplitudes or to use non-Hermitian
Hamiltonians.

\section{Signal-state quantum mechanics}

Regardless of how wave-functions are interpreted, all quantum experiments
are exercises in information extraction. It has been argued above that this
cannot be done in a truly continuous way, even in those experiments designed
to demonstrate the quantum Zeno effect (Bollinger 1990) or support the
predictions of decoherence theory (Zurek 2002). To model this fact, the
formalism of SSQM works with a discrete concept of time rather than a
continuum. What an observer says about their apparatus at any given discrete
time will be labelled by an integer, with a value $n+1$ denoting a stage
later than that denoted by $n$. Each stage represents one of two things:
either a definite change in information about the state of the apparatus, or
else a change in the observer's quantum mechanically based prediction as to
what the state of the apparatus should be at that time, if an observation
were to be made. What these discrete times actually mean in terms of
intervals of physical units of time can be decided later. Moreover, there is
no need to regarded these intervals as all equal.

The elementary detectors modelled in SSQM should not be regarded as
necessarily localized in physical space, or as giving information about
particle position only, although that is certainly possible and may be
important in many situations. These detectors could be constructed from
widely spaced physical components distributed around a laboratory, and
associated with non-localized variables such as particle momenta, or with
spin, or any other physically meaningful concept. What they all have in
common is that they give only \emph{yes} or \emph{no} answers to the
elementary questions $``$has this detector fired or not?$\textquotedblright $

A detector is perhaps best regarded as a physical process which can have one
of two possible outcomes. Given this, it is natural to model such detectors
by \emph{bits} in classical physics and by quantum bits (qubits) in the
quantum scenarios of interest here. The $i^{th}$ detector at time $n$ will
be modelled by the quantum bit $\mathcal{Q}_{n}^{i}$, which is a two
dimensional Hilbert space. Such a qubit should not be identified with any
dichotomous variable such as spin half angular momentum, and is neither a
fermion nor a boson (Wu and Lidar 2002), although the mathematical formalism
suggests the former occasionally.

At any given time $n$, the number $r_{n}$ of detector qubits available to
the observer \emph{at that time }will be called the \emph{rank} of the
corresponding Heisenberg net $\mathcal{H}_{n}\equiv \mathcal{Q}%
_{n}^{1}\otimes \mathcal{Q}_{n}^{2}\otimes \ldots \otimes \mathcal{Q}%
_{n}^{r_{n}}$. This net is the tensor product of all the relevant qubits and
is a Hilbert space of dimension $d_{n}\equiv 2^{r_{n}}$. In those scenarios
involving large numbers of detectors, the associated Heisenberg nets will
have much greater dimensions than the corresponding Hilbert space in SQM,
but there is a sensible physical interpretation of this (Jaroszkiewicz and
Ridgway-Taylor 2006).

Equally significantly, Heisenberg nets contain entangled states as well as
separable states, which is a fundamental feature of quantum mechanics. In
contrast to SQM, the SSQM view is that it is not system states which may be
entangled but states of the apparatus (the labstates). This does not mean
that the apparatus itself is entangled. In conventional experiments,
experimentalists generally know what their apparatus is \footnote{%
SSQM\ has the scope to discuss those situations where an observer is
uncertain about the apparatus itself, rather than about the labstate. The
latter possibility would be represented by a mixed labstate, corresponding
to a mixed state in SQM. The former possibility has no conventional analogue
in SQM. There is no room to comment further on this point here.}. What they
do not know beforehand is how it will behave in quantum outcome terms.

Regarding entanglement as something to do with states of apparatus seems
less objectionable and more natural than thinking of particles themselves as
being entangled. From this point of view, entanglement is just a
manifestation of how the \emph{context} (which is a form of information ) of
an experiment influences a quantum process. For example, abstract four
dimensional Hilbert space and the tensor product of two qubits are
mathematically isomorphic spaces in many respects, but only the latter space
contains entangled elements.

In SSQM there is a natural, preferred basis $B_{n}$ for $\mathcal{H}_{n}$,
consisting of all possible classical (i.e., sharp) signal states. These may
be defined in terms of excitations of the \emph{void state }$|0,n$), the
unique state in $\mathcal{H}_{n}$ for which every detector is in its $``$no$%
\textquotedblright $ state, i.e., $|0,n)\equiv |0,n)_{1}|0,n)_{2}\ldots
|0,n)_{r_{n}}$. Here and elsewhere the tensor product $\otimes $ symbol will
be suppressed. The origin of this uniqueness arises from the knowledge that
the observer has about their apparatus. A Heisenberg net is not just a
tensor product space, but includes the observer's knowledge of what their
apparatus means, and it is this which selects $B_{n}$.

To describe the preferred basis signal states, it is convenient to introduce
a set \{$\mathbb{A}_{i,n}^{+}$, $i=1,2,\ldots ,r_{n}\}$ of \emph{signal
operators} (Jaroszkiewicz and Eakins 2006). There is one signal operator $%
\mathbb{A}_{i,n}^{+}$ associated with each qubit $\mathcal{Q}_{n}^{i}$ in
the Heisenberg net $\mathcal{H}_{n}$, with the property that it changes the
void state $|0,n)_{i}$ in $\mathcal{Q}_{n}^{i}$ to the signal state $%
|1,n)_{i}$ and leaves all the other qubits unaffected, i.e.,
\begin{equation}
\mathbb{A}_{i,n}^{+}|0,n)=|0,n)_{1}\ldots
|0,n)_{i-1}|1,n)_{i}|0,n)_{i+1}\ldots |0,n)_{r_{n}},  \label{333}
\end{equation}%
for $i=1,2,\ldots ,r_{n}$. These signal operators are neither bosonic nor
fermionic. They obey the rules
\begin{equation}
\lbrack \mathbb{A}_{i,n}^{+},\mathbb{A}_{j,n}^{+}]=0,\ \ \ \{\mathbb{A}%
_{i,n},\mathbb{A}_{i,n}^{+}\}=\mathbb{I}_{n},\ \ \ i=1,2,\ldots ,r_{n},
\end{equation}%
where $\mathbb{I}_{n}$ is the identity operator for $\mathcal{H}_{n}$, and
the nilpotency condition%
\begin{equation}
\mathbb{A}_{i,n}^{+}\mathbb{A}_{i,n}^{+}=0\text{,}
\end{equation}%
reminiscent of fermions. The nilpotency condition arises from the fact that
only one signal can be extracted from a given detector at any given time.

The above rules are not ad hoc but consequences of the basic definition (\ref%
{333}) and the completeness of the signal states. It is possible to
construct variants of the signal operators which behave more like fermionic
operators, following the methodology of Jordan and Wigner (Jordan and Wigner
1928, Bjorken and Drell 1965). Under appropriate circumstances, it should be
possible to use their methods to construct the equivalent of fermionic and
bosonic signal field theories (Eakins and Jaroszkiewicz 2005)

The signal operators allow the construction of various \emph{signal classes}%
, consisting of states created by a given number of distinct signal
operators. The zero-signal class consists of one element only, namely the
void state $|0,n)$. The one-signal class consists of the states of the form $%
\mathbb{A}_{i,n}^{+}|0,n)$, and there are exactly $r_{n}$ such states.
Likewise, the two-signal class consists of all states of the form $\mathbb{A}%
_{i,n}^{+}\mathbb{A}_{j,n}^{+}|0,n)$, and there are $r_{n}!/2!(r_{n}-2)!$
such states, and so on.

There are $r_{n}+1$ distinct signal classes, and altogether they give the $%
2^{r_{n}}$ signal states which form the natural basis $B_{n}$ for $\mathcal{H%
}_{n}$. An arbitrary labstate is a normalized linear combination of any of
these basis states. In the applications to particle decays discussed in this
paper, only one-signal states are needed. A one-signal state can sometimes
describe what would correspond to a multi-particle state in SQM. What
determines the interpretation of a labstate is the contextual information
available to the observer as to what their elementary detectors mean. For
example, the question $``$\emph{has this particle decayed into a fireball
consisting of five hundred particles?"} would require only one qubit, not
five hundred.

\subsection{Dynamics}

It is clear from the approach being taken that SSQM avoids the assumptions
of continuity which M\&S referred to in their analysis. In particular, the
Hilbert spaces involved are finite dimensional, reflecting the empirical
fact that all experimentalist have to bin their data, even when it is
assumed that there is a continuum of outcomes.

In SSQM, dynamics is described in terms of mappings of labstates from one
Heisenberg net to its successor Heisenberg net. A \emph{Born map} is defined
here to be any map from one Hilbert space $\mathcal{H}$ to some other
Hilbert space $\mathcal{H}^{\prime }$ which preserves norm, i.e., if $\Psi $
in $\mathcal{H}$ is mapped into $\mathfrak{B}(\Psi )\equiv \Psi ^{\prime }$
in $\mathcal{H}^{\prime }$ by a Born map $\mathfrak{B}$, then%
\begin{equation}
(\Psi ^{\prime },\Psi ^{\prime })=(\Psi ,\Psi ).
\end{equation}%
Born maps are essential here in order to preserve total probability (hence
the terminology), but unfortunately, this is insufficient for quantum
applications. Born maps are not necessarily linear, as can be seen from the
construction of elementary examples. To go further, it is necessary to
impose linearity.

A \emph{semi-unitary operator}\textbf{\ }is defined to be a linear Born map,
i.e., if $\Psi =\alpha \psi +\beta \phi $ for $\psi $,$\phi $ in $\mathcal{H}
$ and $\alpha ,\beta $ complex, then%
\begin{equation}
U\left( \Psi \right) =\alpha U\left( \psi \right) +\beta U\left( \phi \right)
\end{equation}%
if $U$ is semi-unitary. The imposition of linearity, which is an established
feature of SQM, makes a powerful impact in SSQM. It can be readily shown
that a semi-unitary operator $U$ from $\mathcal{H}$ into $\mathcal{H}%
^{\prime }$ exists if and only if $d\equiv \dim \mathcal{H}\leqslant
d^{\prime }\equiv \dim \mathcal{H}^{\prime }$. Moreover, it can be proved
that semi-unitarity implies that
\begin{equation}
U^{+}U=I,
\end{equation}%
where $I$ is the identity for $\mathcal{H}$. From this, it follows
immediately that a semi-unitary operator preserves inner products and not
just norms, i.e., if $\Psi ^{\prime }\equiv U\Psi $ and $\Phi ^{\prime
}\equiv U\Phi $, where $\Psi $ and $\Phi $ are arbitrary elements of $%
\mathcal{H}$, then%
\begin{equation}
(\Phi ^{\prime },\Psi ^{\prime })=(\Phi ,\Psi )
\end{equation}%
if $U$ is semi-unitary. If in fact $d<d^{\prime }$ then necessarily
\begin{equation}
UU^{+}\neq I^{\prime }\text{,}
\end{equation}%
where $I^{\prime }$ is the identity for $\mathcal{H}^{\prime }$.

In SSQM, the dynamics is given in terms of a sequence of semi-unitary
evolution operators $\mathbb{U}_{n+1,n}$ taking labstates in $\mathcal{H}%
_{n} $ to labstates in $\mathcal{H}_{n+1}$ and so on. Such operators satisfy
the rule%
\begin{equation}
\mathbb{U}_{n+1,n}^{+}\mathbb{U}_{n+1,n}=\mathbb{I}_{n}.
\end{equation}%
Further details are given in (Jaroszkiewicz and Eakins, 2006)

\section{One species decays}

In this section, SSQM is used to describe the quantum physics of an unstable
particle state $X$ which can decay to a state $Y$. At all times total
probability will be manifestly conserved.

Every run of an experiment to observe such a decay process will be assumed
to start at time $t=0$, at which time the observer knows that they have
prepared an $X$ state (to use the language of SQM). In SSQM, this is
represented by the labstate $|\Psi ,0)\equiv \mathbb{A}_{X,0}^{+}|0,0)$,
which is automatically normalized to unity.

By time $1$, the labstate will have changed from $|\Psi ,0)$ to some new
labstate $|\Psi ,1)$ of the form
\begin{equation}
|\Psi ,1)=\alpha \mathbb{A}_{X,1}^{+}|0,1)+\beta \mathbb{A}%
_{Y_{1},1}^{+}|0,1),  \label{111}
\end{equation}%
where the complex numbers $\alpha $ and $\beta $ satisfy the semi-unitarity
rule
\begin{equation}
|\alpha |^{2}+|\beta |^{2}=1.
\end{equation}%
The amplitude $\mathcal{A}(X,1|X,0)$ for the particle not to have decayed by
time $1$ is given by the rule
\begin{equation}
\mathcal{A}(X,1|X,0)\equiv (0,1|\mathbb{A}_{X,1}|\Psi ,1)=\alpha
\end{equation}%
whilst the amplitude $\mathcal{A}(Y,1|X,0)$ for the particle to have made
the transition to state $X$ by time $1$ is given by%
\begin{equation}
\mathcal{A}(Y,1|X,0)\equiv (0,1|\mathbb{A}_{Y_{1},1}|\Psi ,1)=\beta .
\label{666}
\end{equation}%
Total probability is therefore conserved. Note that on the right hand side
of (\ref{666}) the label $Y$ is itself labeled by a subscript, in this case
the number $1$, which is the time at which the amplitude is calculated for.
It will be seen that the time at which a transition occurs is a crucial
feature of the analysis, being directly related to the measurement issues
discussed by M\&S.

Such a process conserves signal class, so the dynamics can be discussed
wholly in terms of the evolution of the signal operators rather than the
labstates. For instance, evolution from $0$ to $1$ can be given in the form%
\begin{equation}
\mathbb{A}_{X,0}^{+}\rightarrow \mathbb{U}_{1,0}\mathbb{A}_{X,0}^{+}\mathbb{U%
}_{1,0}^{+}=\alpha \mathbb{A}_{X,1}^{+}+\beta \mathbb{A}_{Y_{1},1}^{+},
\label{888}
\end{equation}%
where $\mathbb{U}_{1,0}$ is a semi-unitary operator satisfying the rule%
\begin{equation}
\mathbb{U}_{1,0}^{+}\mathbb{U}_{1,0}=\mathbb{I}_{0}\text{,}
\end{equation}%
with $\mathbb{I}_{0}$ being the identity for the initial Heisenberg net $%
\mathcal{H}_{0}\equiv \mathcal{Q}_{0}^{X}.$

The above process involves a change in rank, since $\mathcal{H}_{1}\equiv
\mathcal{Q}_{1}^{X}\otimes \mathcal{Q}_{1}^{Y_{1}}$. Because $r_{1}\equiv
\dim \mathcal{H}_{1}>r_{0}\equiv \dim \mathcal{H}_{0}$, semi-unitarity of
the evolution operator means that%
\begin{equation}
\mathbb{U}_{1,0}\mathbb{U}_{1,0}^{+}\neq \mathbb{I}_{1}\text{.}
\end{equation}%
Unlike SQM, therefore, the SSQM approach to quantum dynamics places clear
constraints on what is meant by time reversal. Any attempt to discuss time
reversal has to focus on the apparatus involved. Moreover, any such
experiment always takes place in the same direction of time as the rest of
the universe.

The description of the next stage of the process, from time $1$ to time $2$,
is more subtle and involves the concept of \emph{null test} . Such a test is
defined here as any quantum test which extracts no information from a given
initial state. In SQM, this corresponds to passing some outcome of an
apparatus through the same or equivalent apparatus, the net effect being to
leave the state unchanged. For example, an electron emerging from a
Stern-Gerlach apparatus $S_{0}$ in the spin-up state will pass through
another Stern-Gerlach apparatus $S_{1}$ completely unscathed, provided the
magnetization axis of $S_{1}$ is in the same direction as that of $S_{0}$.
In SQM, a null test is modelled mathematically by the fact that an
eigenstate of an operator is also an eigenstate of the square of that
operator.

Considering the labstate of the above decay process at time $1$, there are
now two terms to consider. The first term in (\ref{888}), $\alpha \mathbb{A}%
_{X,1}^{+}$, corresponding to \emph{no decay} by time $1$, can be regarded
at this point as creating an initial $X$ state which could now decay into a $%
Y$ state or not, with the same characteristics as for the first stage of the
run, i.e., between times $0$ and $1$. This assumes spatial and temporal
homogeneity, a physically reasonable assumption which would need to be
reconsidered in gravitational fields, which are taken to be absent here for
simplicity. The second term, $\beta \mathbb{A}_{Y_{1},1}^{+}$, corresponds
to \emph{decay having occurred during the first time interval},\emph{\ }%
which is regarded here as irreversible.\emph{\ }Situations where the $Y$
state can revert back to the $X$ state are more complicated and of greater
empirical interest, such in the ammonium maser and neutral Kaon decay, and
these are discussed in later sections of this paper.\emph{\ }

Assuming homogeneity, the next stage of the evolution is given by%
\begin{eqnarray}
\mathbb{U}_{2,1}\mathbb{A}_{X,1}^{+}\mathbb{U}_{2,1}^{+} &=&\alpha \mathbb{A}%
_{X,2}^{+}+\beta \mathbb{A}_{Y_{2},2}^{+},  \notag \\
\mathbb{U}_{2,1}\mathbb{A}_{Y_{1},1}^{+}\mathbb{U}_{2,1}^{+} &=&\mathbb{A}%
_{Y_{1},2}^{+}.  \label{222}
\end{eqnarray}%
The second equation is justified as follows. The decay term in (\ref{888}),
proportional to $\mathbb{A}_{Y_{1},1}^{+}$ at time $1,$ corresponds to the
possibility of detecting a decay product state at that time. Now there is
nothing which requires this information to be extracted precisely at that
time. The experimentalist could choose to delay information extraction until
some later time, effectively placing the decay product observation $``$on
hold$\textquotedblright $. As stated above, this may be represented in SQM\
by passing a state through a null-test, which does not alter it. In SSQM\
this is represented by the second equation in (\ref{222}). Essentially,
quantum information about a decay is passed forwards in time until it is
physically extracted.

The Heisenberg net $\mathcal{H}_{2}$ at time $2$ has rank three, being the
tensor product%
\begin{equation}
\mathcal{H}_{2}=\mathcal{Q}_{2}^{X}\otimes \mathcal{Q}_{2}^{Y_{1}}\otimes
\mathcal{Q}_{2}^{Y_{2}}.
\end{equation}%
Semi-unitary evolution from time zero to time $2$ therefore gives%
\begin{eqnarray}
\mathbb{A}_{X,0}^{+}\rightarrow \mathbb{U}_{2,1}\mathbb{U}_{1,0}\mathbb{A}%
_{X,0}^{+}\mathbb{U}_{1,0}^{+}\mathbb{U}_{2,1}^{+} &=&\alpha ^{2}\mathbb{A}%
_{X,2}^{+}+\alpha \beta \mathbb{A}_{Y_{2},2}^{+}  \notag \\
&&+\beta \mathbb{A}_{Y_{1},2}^{+},
\end{eqnarray}%
with the various probabilities being read off as the squared moduli of the
corresponding terms.

It is clear that in (\ref{123}) a space-time description with a specific
arrow of time is being built up, with a memory of the change of rank of the
Heisenberg net at time $1$ being propagated forwards in time to time $2.$
This is represented by the contribution involving $\mathbb{A}_{Y_{1},2}^{+},$
from a potential decay process which may have occurred by time $1$, and this
contributes to the overall labstate amplitude at time $2$.

Subsequently the process continues in an analogous fashion, with the rank of
the Heisenberg net increasing by one at each timestep. At time $n$ the
dynamics gives%
\begin{equation}
\mathbb{A}_{X,0}^{+}\rightarrow \mathbb{U}_{n,0}\mathbb{A}_{X,0}^{+}\mathbb{U%
}_{n,0}^{+}=\alpha ^{n}\mathbb{A}_{X,n}^{+}+\beta \sum_{k=1}^{n}\alpha ^{k-1}%
\mathbb{A}_{Y_{k},n}^{+},
\end{equation}%
where $\mathbb{U}_{n,0}\equiv \mathbb{U}_{n,n-1}\mathbb{U}_{n-1,n-2}\ldots
\mathbb{U}_{1,0}$ is semi-unitary and satisfies the constraint%
\begin{equation}
\mathbb{U}_{n,0}^{+}\mathbb{U}_{n,0}=\mathbb{I}_{0}.
\end{equation}%
From the above, the survival probability $\Pr (X,n|X,0)$ that the original
state has \emph{not} decayed can be easily read off and is found to be%
\begin{equation}
\Pr (X,n|X,0)=|\alpha |^{2n}.
\end{equation}%
\qquad Provided $\beta \neq 0$, this probability appears to falls
monotonically with increasing $n$, which corresponds to particle decay.

The discussion at this point calls for some care with limits, because here
there arises the theoretical possibility of encountering the quantum Zeno
effect, as discussed by M\&S. In the following, it will be assumed that $%
|\alpha |<1$, because $|\alpha |=1$ corresponds to a stable particle, which
is of no interest here.

Now consider the physics of the situation. The calculated probabilities
should relate to the elapsed time $t$ as used by the observer in the
laboratory, which is not assumed here to be a continuous parameter. The
temporal subscript $n$ labelling successive stages corresponds to an elapsed
time given by $t\equiv n\tau $, where $\tau $ is some well-defined time
scale characteristic of the apparatus. In the sort of experiments relevant
here, $\tau $ will be a very small fraction of a second, but certainly
nowhere near the Planck time. Realistic measurement timescales, involving
electromagnetic processes, would be in the $10^{-9}-10^{-18}$ second range.

If now the transition amplitude $\alpha $ and $\tau $ are related by the
rule
\begin{equation}
|\alpha |^{2}\equiv e^{-\Gamma \tau },  \label{999}
\end{equation}%
where $\Gamma $ is a characteristic inverse time, then the survival
probability $P\left( t_{n}\right) $ is given by%
\begin{equation}
P\left( t\right) \equiv \Pr (X,n|X,0)=e^{-\Gamma t},
\end{equation}%
which is the usual exponential decay formula. No imaginary term proportional
to $\Gamma $ in any supposed Hamiltonian or energy has been introduced in
order to obtain this exponential decay law.

A subtlety arise here however. Expression (\ref{999}) is equivalent to the
assumption that $|\alpha |^{2}$ is an analytic function of $\tau $ with a
Taylor expansion of the form%
\begin{equation}
|\alpha |^{2}=1-\Gamma \tau +O\left( \tau \right) ^{2},  \label{444}
\end{equation}%
i.e., one with a non-zero linear term. Under such circumstances, the
standard result $\lim_{n\rightarrow \infty }(1-\frac{x}{n})^{n}=e^{-x}$
gives the exponential decay law. The possibility remains, however, that the
dynamics is such that the linear term is zero, so that the appropriate
expression is of the form%
\begin{equation}
|\alpha |^{2}=1-\gamma \tau ^{2}+O\left( \tau ^{3}\right) ,  \label{555}
\end{equation}%
where $\gamma $ is a positive constant (Bollinger et al. 1990). Then in the
limit $n\rightarrow \infty $, where $n\tau \equiv t$ is held fixed, the
result is given by%
\begin{equation}
\lim_{n\rightarrow \infty ,n\tau =t\ \text{fixed}}\left( 1-\gamma \tau
^{2}+O\left( \tau ^{3}\right) \right) ^{n}=1,
\end{equation}%
which gives rise to the quantum Zeno effect scenario.

To understand properly what is going on, it is necessary to appreciate that
there are two competing limits being considered: one where a system is being
observed over an increasing macroscopic laboratory time scale $t$, and
another one where many observations are being taken in succession, separated
by a microscopic time scale $\tau $ which is being brought as close to zero
as possible. In each case, the limit cannot be achieved in the laboratory.
The result is that in such experiments, the apparatus may play a decisive
role in determining the measured outcomes. If the apparatus is such that (%
\ref{444}) holds, then exponential decay will be observed. If on the other
hand the apparatus behaves according to the rule (\ref{555}), or any
reasonable variant of it, then approximations to the quantum Zeno effect
should be observed.

The scenario where apparatus plays a decisive role in observation was
discussed by M\&S, but not taken further, on the grounds that they saw no
indication that the $``"$\emph{observed lifetime}$"$ \emph{of an unstable
particle is not a property of the particle (and its Hamiltonian)}$%
\textquotedblright $\emph{\ }(Misra and Sudarshan 1977). It is probable that
M\&S included a reference a Hamiltonian because they recognized that,
contrary to what they wanted to assert (i.e., that the dynamical evolution
of a decaying particle is an intrinsic property of the particle alone), the
external environment does play a role in observation. In fact, Hamiltonians
change whenever the environment in which particles are situated is changed
by the observer. For instance, if an electric field is switched on, the
Hamiltonian associated with a charged particle changes. It is therefore
incorrect to regard a Hamiltonian as an intrinsic property of a system under
observation alone.

Our conclusion is that there is \emph{every} indication that apparatus plays
a role in the observed dynamics of particles.

\subsection{Semi-unitary matrices}

The above scenario can be discussed more efficiently in terms of
semi-unitary matrices, an approach which will be useful in the discussion of
neutral Kaon decay given later.

A semi-unitary matrix $\mathsf{M}$ is a $r^{\prime }\times r$ matrix mapping
complex $r-$dimensional column vectors into complex $r^{\prime }-$%
dimensional column vectors$,$ such that%
\begin{equation}
\mathsf{M}^{+}\mathsf{M}=\mathsf{I}_{r},
\end{equation}%
where $\mathsf{I}_{r}$ is the $r\times r$ identity matrix. No semi-unitary
matrix exists if $r^{\prime }<r$.

Now consider the $X$ decay scenario discussed above. If the initial labstate
is represented by the $1\times 1$ matrix $\Psi _{0}\equiv \lbrack 1]$, then
the action of $\mathbb{U}_{1,0}$ given by (\ref{111}) may be represented by
the semi-unitary matrix
\begin{equation}
\mathsf{U}_{1,0}\equiv \left[
\begin{array}{l}
\alpha \\
\beta%
\end{array}%
\right] .
\end{equation}%
Then the labstate at time $1$ is represented by the $2\times 1$ matrix $\Psi
_{1}$ given by
\begin{equation}
\Psi _{1}=\mathsf{U}_{1,0}\Psi _{0}=\left[
\begin{array}{l}
\alpha \\
\beta%
\end{array}%
\right] [1]=\left[
\begin{array}{l}
\alpha \\
\beta%
\end{array}%
\right] .
\end{equation}%
The two required transition amplitudes are just the various components of
this vector.

Going further, the action of $\mathbb{U}_{2,0}$ is represented by the $%
3\times 2$ semi-unitary matrix%
\begin{equation}
\mathsf{U}_{2,1}=\left[
\begin{array}{ll}
\alpha & 0 \\
\beta & 0 \\
0 & 1%
\end{array}%
\right] ,
\end{equation}%
and so on for later times. For arbitrary $n>1$, it is found that%
\begin{equation}
\mathsf{U}_{n,n-1}=\left[
\begin{array}{ll}
\alpha & 0 \\
\beta & 0 \\
0 & \mathsf{I}_{n-1}%
\end{array}%
\right] ,
\end{equation}%
where $\mathsf{I}_{n}$ is the $n\times n$ identity matrix. This gives%
\begin{equation}
\Psi _{n}=\mathsf{U}_{n,n-1}\mathsf{U}_{n-1,n-2}\ldots \mathsf{U}_{1,0}\Psi
_{0}=\left[
\begin{array}{c}
\alpha ^{n} \\
\beta \alpha ^{n-1} \\
\vdots \\
\beta \alpha \\
\beta%
\end{array}%
\right] .
\end{equation}

The squared modulus of the first component of this column vector gives the
same survival probability $|\alpha |^{2n}$ as before. It is also relatively
easy now to read off all the other probabilities and give a discrete time
version of the $P$, $Q$ and $R$ functions discussed by M\&S. Let $P_{n}$ be
the probability that an $X$ state created at time $0$ has made a transition
to a $Y$ state sometime during the temporal interval $[1,n],$ inclusive of
the end times, let $Q_{n}$ be the probability that an initially prepared $X$
state has not made such a transition at any time during this interval, and
let $R_{m,n}$ be the probability that an initially prepared $X$ state had
not made a transition between time $0$ and $m$, and then made a transition
sometime between $m+1$ and $n$ inclusive (assuming $0\leqslant m<n).$ Then
clearly%
\begin{equation}
P_{n}+Q_{n}=0,\ \ \ R_{m,n}=Q_{m}P_{n-m}.
\end{equation}%
From the components of the $\Psi _{n}$, it is found that%
\begin{equation}
P_{n}=1-|\alpha |^{2n},\ \ \ Q_{n}=|\alpha |^{2n},\ \ \ n=0,1,2,\ldots ,
\end{equation}%
from which $R_{m,n}$ can be constructed.

Although this analysis gives results which look formally like the standard
decay result, the scenario is equivalent to that discussed by M\&S, namely,
there is a constant questioning (or its discrete equivalent) by the
apparatus as to whether decay has taken place or not. In this case the
results are simple. In more complicated scenarios, such as Kaon decay, the
results are more complicated.

\section{The Ammonium system}

A successful application of SQM to particle physics was the explanation by
Gell-Mann and Pais (Gell-Mann and Pais 1955) of the phenomenon of
regeneration in neutral Kaon decays. In the standard calculation (Leighton,
Feynman, and Sands 1966), a non-hermitian Hamiltonian is used to introduce
the two decay parameters needed to describe the observations. In the next
section it will be shown how SSQM readily reproduces the result of the
Gell-Mann-Pais calculation whilst conserving total probability.

The analysis of the Kaon system is more complex than the single particle
decay process discussed above, involving the interplay of two distinct
neutral Kaons, the $K^{0}$ and its antiparticle, the $\bar{K}^{0}.$ In order
to understand the SSQM approach to the description of neutral Kaon decays,
it will be helpful to review first how systems such as the ammonium molecule
are discussed in SQM and in SSQM.

When translation and rotational symmetries are ignored, the ammonium
molecule is described in SQM in terms of a superposition of two orthonormal
states representing the two possible position states of the single nitrogen
atom relative to the plane defined by the three hydrogen atoms. These two
states form a basis for a two-dimensional Hilbert space describing the
system. The Hamiltonian for the system is represented by the Hermitian matrix%
\begin{equation}
\mathsf{H}=\left[
\begin{array}{ll}
e & f \\
f^{\ast } & g%
\end{array}%
\right] ,
\end{equation}%
where $e$ and $g$ are real and $f$ can be complex. If the state of the
molecule is represented at time $t$ by the two-component wave-function
\begin{equation}
\Psi (t)\equiv \left[
\begin{array}{l}
\Psi _{1}(t) \\
\Psi _{2}(t)%
\end{array}%
\right] ,
\end{equation}%
then the Schr\"{o}dinger equation $i\hbar \partial _{t}\Psi (t)=\mathsf{H}%
\Psi (t)$ has solutions%
\begin{equation}
\Psi _{j}\left( t\right) =A_{j}e^{-i\omega ^{+}t}+B_{j}e^{-i\omega ^{-}t},\
\ \ j=1,2,
\end{equation}%
where $A_{j}$ and $B_{j}$ are constants and $\omega ^{\pm }=\frac{1}{2}%
\{e+g\pm \sqrt{4|f|^{2}+(e-g)^{2}}\}.$ This gives probability functions
which have oscillatory behaviour with a frequency given by the difference $%
\omega ^{+}-\omega ^{-}$.

In the SSQM description, it will be assumed that there are two different
states, $X$, $Y$, with signal operators $\mathbb{A}_{X,n}^{+}$, $\mathbb{A}%
_{Y,n}^{+}$ respectively, evolving according to the rule%
\begin{eqnarray}
\mathbb{U}_{n+1,n}\mathbb{A}_{X,n}^{+}|0,n) &=&\{a\mathbb{A}_{X,n+1}^{+}+b%
\mathbb{A}_{Y,n+1}^{+}\}|0,n+1),  \notag \\
\mathbb{U}_{n+1,n}\mathbb{A}_{Y,n}^{+}|0,n) &=&\{c\mathbb{A}_{X,n+1}^{+}+d%
\mathbb{A}_{Y,n+1}^{+}\}|0,n+1),  \notag \\
&&
\end{eqnarray}%
where $\mathbb{U}_{n+1,n}$ is a semi-unitary operator. Semi-unitarity
requires the constraints%
\begin{equation}
|a|^{2}+|b|^{2}=|c|^{2}+|d|^{2}=1,\ \ \ a^{\ast }c+b^{\ast }d=0.
\end{equation}%
All other states will be disregarded on the basis that there are no
dynamical channels between them and states $X$ and $Y$. With a suitable
choice of phases, $\mathbb{U}_{n+1,n}$ can be represented by the
semi-unitary matrix%
\begin{equation}
\mathsf{U}=\left[
\begin{array}{rr}
a & -b^{\ast } \\
b & a^{\ast }%
\end{array}%
\right] ,
\end{equation}%
where an overall possible phase factor is ignored and $a$ and $b$ are as
above. The eigenvalues $z^{\pm \text{ }}$of $\mathsf{U}$ are given by%
\begin{equation}
z^{\pm }=%
{\frac12}%
\left\{ a+a^{\ast }\pm i\sqrt{4-(a+a^{\ast })^{2}}\right\} .
\end{equation}%
These are complex conjugates of each other and have magnitude unity, so can
be written in the form $z^{\pm }=\exp \left\{ \pm i\theta \right\} ,$where $%
\theta $ is real. Writing $a\equiv |a|e^{i\alpha },$ where $\alpha $ is
real, then $\cos \theta =|a|\cos \alpha $. Now $\mathsf{U}$ can always be
written in the form%
\begin{equation}
\mathsf{U}=\mathsf{V}\left[
\begin{array}{cc}
e^{i\theta } & 0 \\
0 & e^{-i\theta }%
\end{array}%
\right] \mathsf{V}^{+},
\end{equation}%
where $\mathsf{V}$ is semi-unitary. Again, with a suitable choice of phases,
$\mathsf{V}$ can be written in the form%
\begin{equation}
\mathsf{V}=\left[
\begin{array}{cc}
u & -v^{\ast } \\
v & u^{\ast }%
\end{array}%
\right] ,
\end{equation}%
where $|u|^{2}+|v|^{2}=1$, and then it is found that%
\begin{equation}
|u|^{2}e^{i\theta }+|v|^{2}e^{-i\theta }=a,\ \ \ u^{\ast }v(e^{i\theta
}-e^{-i\theta })=b.
\end{equation}%
This gives%
\begin{eqnarray}
\mathsf{U}^{n} &=&\mathsf{V}\left[
\begin{array}{cc}
e^{in\theta } & 0 \\
0 & e^{-in\theta }%
\end{array}%
\right] \mathsf{V}^{+}  \notag \\
&=&\left[
\begin{array}{cc}
|u|^{2}e^{in\theta }+|v|^{2}e^{-in\theta } & uv^{\ast }\{e^{in\theta
}-e^{-in\theta }\} \\
u^{\ast }v\{e^{in\theta }-e^{-in\theta }\} & |u|^{2}e^{-in\theta
}+|v|^{2}e^{in\theta }%
\end{array}%
\right] ,  \notag \\
&&
\end{eqnarray}%
which leads to the dynamical rule
\begin{eqnarray}
\mathbb{A}_{X,0}^{+} &\rightarrow &\{|u|^{2}e^{in\theta
}+|v|^{2}e^{-in\theta }\}\mathbb{A}_{X,n}^{+}  \notag \\
&&\ \ \ \ \ \ \ \ \ \ +u^{\ast }v\{e^{in\theta }-e^{-in\theta }\}\mathbb{A}%
_{Y,n}^{+},  \notag \\
\mathbb{A}_{Y,0}^{+} &\rightarrow &uv^{\ast }\{e^{in\theta }-e^{-in\theta }\}%
\mathbb{A}_{X,n}^{+}  \notag \\
&&\ \ \ \ \ \ \ \ \ \ +\left\{ |u|^{2}e^{-in\theta }+|v|^{2}e^{in\theta
}\right\} \mathbb{A}_{Y,n}^{+}.
\end{eqnarray}%
Hence the probabilities are given by%
\begin{eqnarray}
\Pr \left( X,n|X,0\right) &=&\Pr (Y,n|Y,0)  \notag \\
&=&|u|^{4}+|v|^{4}+2|u|^{2}|v|^{2}\cos \left( 2n\theta \right) ,  \notag \\
\Pr (Y,n|X,0) &=&\Pr (X,n|Y,0)  \notag \\
&=&4|u|^{2}|v|^{2}\sin ^{2}\left( n\theta \right) ,
\end{eqnarray}%
which agrees with the SQM expressions when $2n\theta =(\omega ^{+}-\omega
^{-})t.$

It was noted in (Bollinger et al. 1990) that a survival probability of the
form $P(\tau )\ \sim \ 1-\gamma \tau ^{2}+O(\tau ^{3})$ would be needed to
make observations of the quantum Zeno effect viable. The above calculation
of the ammonium survival probabilities is compatible with this, as can be
seen from the expansion%
\begin{eqnarray}
\Pr \left( X,n|X,0\right) &=&|u|^{4}+|v|^{4}+2|u|^{2}|v|^{2}\cos \left(
2n\theta \right) \   \notag \\
&\sim &1-4|u|^{2}|v|^{2}n^{2}\theta ^{2}+O\left( n^{4}\theta ^{4}\right) .
\end{eqnarray}%
Therefore, it is expected that the quantum Zeno effect (or at least some
behaviour analogous to it) should be observable in the ammonium system,
provided the amplitudes involved have the required dependence on $\tau $. As
with the particle decays discussed above, it would be necessary to ensure
that the two limits, $n\rightarrow \infty $, $\tau \rightarrow 0$, were
carefully balanced, the point being that the laboratory protocol will play a
decisive role in the outcome of the experiment.

\section{Kaon-type decays}

More complex systems such as neutral Kaon decay are readily discussed in
SSQM as follows. Consider three different particle states, $X,Y$ and $Z,$
making transitions between each other in the specific way described below.
An important example occurring in particle physics involves the neutral
Kaons, with $X$ representing a $K^{0}$ meson, $Y$ representing a $\bar{K}%
^{0} $ meson, and $Z$ representing their various decay products. Kaon decay
is remarkable for the phenomenon of regeneration, in which the Kaon survival
probabilities fall and then rise with time. More recently, a similar
phenomenon has been observed in $B$ meson decay.

As before, attention can be focused on one-signal states states. The
dynamics is described by the transition rules%
\begin{eqnarray}
\mathbb{A}_{X,n}^{+} &\rightarrow &\alpha \mathbb{A}_{X,n+1}^{+}+\beta
\mathbb{A}_{Y,n+1}^{+}+\gamma \mathbb{A}_{X_{n+1},n+1}^{+},  \notag \\
\mathbb{A}_{Y,n}^{+} &\rightarrow &u\mathbb{A}_{X,n+1}^{+}+v\mathbb{A}%
_{Y,n+1}^{+}+w\mathbb{A}_{Z_{n+1},n+1}^{+},  \notag \\
\mathbb{A}_{Z_{n},n}^{+} &\rightarrow &A_{Z_{n},n+1}^{+},  \label{777}
\end{eqnarray}%
where semi-unitarity requires the transition coefficients to satisfy the
constraints%
\begin{eqnarray}
|\alpha |^{2}+|\beta |^{2}+\gamma |^{2} &=&1,  \notag \\
|u|^{2}+|v|^{2}+|w|^{2} &=&1, \\
\alpha ^{\ast }u+\beta ^{\ast }v+\gamma ^{\ast }w &=&0.  \notag
\end{eqnarray}%
The above process is a combination of the decay and oscillation processes
discussed previously.

By inspection of (\ref{777}), it can be seen that there are no transitions
from $Z$ states to either $X$ or $Y$ states. Therefore, once a $Z$ state is
created, it remains a $Z$ state. There is an irreversible flow from the $X$
and $Y$ states, so these eventually disappear. Before that occurs however,
there will be back-and-forth transitions between the $X$ and $Y$ states
which give rise to the phenomenon of regeneration.

In actual Kaon decay experiments, pure $K^{0}$ states can be prepared via
the strong interaction process $\pi ^{-}+p\rightarrow K^{0}+\Lambda ,$
whilst pure $\bar{K}^{0}$ states can be prepared via the process $\pi
^{+}+p\rightarrow K^{+}+\bar{K}^{0}+p.$ In our notation, these preparations
correspond to initial labstates\ $\mathbb{A}_{X,0}^{+}|0,0)$ and $\mathbb{A}%
_{Y,0}^{+}|0,0)$ respectively. In practice, superpositions of $K^{0}$ and $%
\bar{K}^{0}$ states may be difficult to prepare directly, but the analysis
of Gell-Mann and Pais shows that such states can be obtained indirectly
(Gell-Mann and Pais 1955). Therefore, labstates corresponding to $X$ and $Y$
superpositions are physically meaningful and will be used in the following
analysis.

Consider an initial labstate of the form%
\begin{equation}
|\Psi ,0)\equiv \left\{ a\mathbb{A}_{X,0}^{+}+b\mathbb{A}_{Y,0}^{+}\right\}
|0,0),
\end{equation}%
where $|a|^{2}+|b|^{2}=1$. Matrix methods are appropriate here. The dynamics
of the system will be discussed in terms of the initial column vector%
\begin{equation}
\Psi _{0}\equiv \left[
\begin{array}{c}
a \\
b%
\end{array}%
\right] ,
\end{equation}%
which is equivalent to the statement that each run of the experiment starts
with the rank-two Heisenberg net $\mathcal{H}_{0}\equiv \mathcal{Q}%
_{0}^{X}\otimes \mathcal{Q}_{0}^{Y}$. The dynamical rules (\ref{777}) map
labstates in $\mathcal{H}_{0}$ into $\mathcal{H}_{1}\equiv \mathcal{Q}%
_{1}^{X}\otimes \mathcal{Q}_{1}^{Y}\otimes \mathcal{Q}_{1}^{Z_{1}}$, so
there is a change of rank from two to three. The transition is represented
by the semi-unitary matrix%
\begin{equation}
\mathsf{U}_{1,0}\equiv \left[
\begin{array}{cc}
\alpha & u \\
\beta & v \\
\gamma & w%
\end{array}%
\right] .
\end{equation}%
More generally, we may write%
\begin{equation}
\mathsf{U}_{n+1,n}\equiv \left[
\begin{array}{ccc}
\alpha & u & 0 \\
\beta & v & 0 \\
\gamma & w & 0 \\
0 & 0 & \mathsf{I}_{n}%
\end{array}%
\right] ,\ \ \ n>0,
\end{equation}%
where $\mathsf{I}_{n}$ is the $n\times n$ identity. The Heisenberg net at
time $n$ has rank $n+2$ and changes to one of rank $n+3$ over the next time
step.

If the state at time $t$ is represented by a column vector $\Psi _{n}$ with $%
n+2$ components, then we may write%
\begin{equation}
\Psi _{n}=\mathsf{U}_{n,n-1}\mathsf{U}_{n-1,n-2}\ldots \mathsf{U}_{2,1}%
\mathsf{U}_{2,0}\Psi _{0}.
\end{equation}%
Overall probability is conserved, because the semi-unitarity of the
transition operators $\mathsf{U}_{i+1,i}$ guarantees that
\begin{equation}
\Psi _{n}^{+}\Psi _{n}=\Psi _{0}^{+}\Psi _{0}.
\end{equation}

Once again, the key to unravelling the dynamics is to use linearity, which
is guaranteed by the use of semi-unitary evolution operators. Suppose the
state $\Psi _{n}$ at time $n$ is represented by
\begin{equation}
\Psi _{n}=\left[
\begin{array}{c}
x_{n} \\
y_{n} \\
z_{n,n} \\
\vdots \\
z_{1,n}%
\end{array}%
\right] ,
\end{equation}%
where the components $x_{n}$ and $y_{n}$ are such that%
\begin{equation}
x_{n}=\lambda ^{n}x_{0},\text{ }y_{n}=\lambda ^{n}y_{0}\text{,}
\end{equation}%
where $\lambda $ is some complex number to be determined. Such states will
be referred to as eigenmodes. They are not eigenstates of any physical
operator, but their first two components, $x_{n}$ and $y_{n}$ behave as if
they were. Then the dynamics gives the following relations:%
\begin{eqnarray}
x_{n+1} &=&\alpha x_{n}+y_{n}=\lambda x_{n},  \notag \\
x_{n+1} &=&\beta x_{n}+vy_{n}=\lambda y_{n},  \notag \\
z_{n+1,n+1} &=&\gamma x_{n}+wy_{n}.
\end{eqnarray}%
Experimentalists will be interested only in survival probabilities for the $%
X $ and $Y$ states, so the dynamics of $Z$ states will be ignored here,
i.e., the behaviour of the components $z_{k,n}$ for $k<n$ will not be
discussed$.$

It will be seen from the above that $\lambda $ is an eigenvalue of the matrix%
\begin{equation}
\mathsf{M}\equiv \left[
\begin{array}{cc}
\alpha & u \\
\beta & v%
\end{array}%
\right] ,
\end{equation}%
which means that in principle there are two solutions, $\lambda ^{+}$ and $%
\lambda ^{-}$, for the eigenmode values, given by%
\begin{equation}
\lambda ^{\pm }=\frac{\alpha +v\pm \sqrt{(\alpha -v)^{2}+4\beta u}}{2}.
\end{equation}%
It is expected that these will not be mutual complex conjugates in actual
experiments, because if they were, the analysis could not explain observed
Kaon physics. Therefore, the coefficients $\alpha $,$\beta ,u$ and $v$ will
be such that the above two eigenmode values are complex and of different
magnitude and phase, giving rise to two decay channels with different
lifetimes, as happens in neutral Kaon decay. In the SQM analysis of neutral
Kaon decays, Gell-Mann and Pais described the neutral Kaons as
superpositions of two hypothetical particles known as $K_{1}^{0}$ and $%
K_{2}^{0}$, which are $CP$ eigenstates and have different decay lifetimes
(Gell-Mann and Pais 1955). The $K_{1}^{0}$ decays to a two pion state with a
lifetime of about $0.9\times 10^{-10}$ seconds whilst the $K_{2}^{0}$ decays
to a three pion state with a lifetime of about $0.5\times 10^{-7}$ seconds.

Semi-unitarity guarantees that%
\begin{equation}
|x_{n+1}|^{2}+|y_{n+1}|^{2}+|z_{n+1,n+1}|^{2}=|x_{n}|^{2}+|y_{n}|^{2},
\end{equation}%
and so it can be deduced that%
\begin{equation}
|\lambda |^{2}=1-\frac{|z_{n+1,n+1}|^{2}}{|x_{n}|^{2}+|y_{n}|^{2}}<1,\ \ \ \
\ n=0,1,2,\ldots
\end{equation}%
From this and the conditions%
\begin{equation}
x_{n}=\lambda ^{n}x_{0},\ \ \ y_{n}=\lambda ^{n}y_{0},
\end{equation}%
the eigenmode values can be written in the form%
\begin{equation}
\lambda _{1}=r_{1}e^{i\theta _{1}},\ \ \ \lambda _{1}=r_{2}e^{i\theta _{2}},
\end{equation}%
where $r_{1}<1$, $r_{2}<1$ and $\theta _{1}$ and $\theta _{2}$ are real. The
eigenmodes at time $t=0$ corresponding to $\lambda _{1}$ and $\lambda _{2}$
will be denoted by $\Lambda _{1,0}$ and $\Lambda _{2,0}$ respectively, i.e.%
\begin{equation}
\Lambda _{1,0}=\left[
\begin{array}{c}
a_{1} \\
b_{1}%
\end{array}%
\right] \text{, }\Lambda _{2,0}=\left[
\begin{array}{c}
a_{2} \\
b_{2}%
\end{array}%
\right] \text{,}
\end{equation}%
and then the evolution rules give%
\begin{equation}
\Lambda _{1,n}=\left[
\begin{array}{c}
\lambda _{1}^{n}a_{1} \\
\lambda _{1}^{n}b_{1} \\
c_{n,n} \\
\vdots \\
c_{1,n}%
\end{array}%
\right] ,\ \ \ \Lambda _{2,n}=\left[
\begin{array}{c}
\lambda _{2}^{n}a_{2} \\
\lambda _{1}^{n}b_{2} \\
d_{n,n} \\
\vdots \\
d_{1,n}%
\end{array}%
\right] ,
\end{equation}%
where the coefficients $\left\{ c_{k,n}\right\} ,\left\{ d_{k,n}\right\} $
can be determined from the dynamics. The initial modes $\Lambda _{1,0}$ and $%
\Lambda _{2,0}$ are linearly independent provided $\lambda _{1}$ and $%
\lambda _{2}$ are different. Given that, then any initial labstate $\Psi
_{0} $ can be expressed uniquely as a normalized linear combination of $%
\Lambda _{1,0}$ and $\Lambda _{2,0},$ i.e.,
\begin{equation}
\Psi _{0}=\mu _{1}\Lambda _{1,0}+\mu _{2}\Lambda _{2,0},
\end{equation}%
for some coefficients $\mu _{1}$ and $\mu _{2}$. This is the analogue of the
decompositions%
\begin{equation}
|K^{0}\rangle =\frac{1}{\sqrt{2}}\left\{ |K_{1}^{0}\rangle
+|K_{2}^{0}\rangle \right\} \text{,}\ \ \ |\bar{K}^{0}\rangle =\frac{1}{%
\sqrt{2}}\left\{ |K_{1}^{0}\rangle -|K_{2}^{0}\rangle \right\}
\end{equation}%
in the Gell-Mann-Pais approach.

From this, the amplitude $\mathcal{A}(X,n|\Psi ,0)$ to find an $X$ signal at
time $n$ is given by%
\begin{equation}
\mathcal{A}\left( X,n|\Psi ,0\right) =\mu _{1}a_{1}\lambda _{1}^{n}+\mu
_{2}a_{2}\lambda _{2}^{n},
\end{equation}%
so that the survival probability for $X$ is given by%
\begin{eqnarray}
\Pr \left( X,n|\Psi ,0\right) &=&|\mu _{1}|^{2}|a_{1}|^{2}r_{1}^{2n}+|\mu
_{2}|^{2}|a_{2}|^{2}r_{2}^{2n}  \notag \\
&&+2r_{1}^{n}r_{2}^{n}\func{Re}\left\{ \mu _{1}^{\ast }\mu _{2}a_{1}^{\ast
}a_{2}e^{-in(\theta _{1}-\theta _{2})}\right\} ,  \notag \\
&&
\end{eqnarray}%
and similarly for $\Pr \left( Y,n|\Psi ,0\right) $.

There is scope here for various limits to be considered, as discussed in the
single channel decay analysis, such that either particle decay is seen or
the quantum Zeno effect appears to hold over limited time spans. If we write
\begin{equation}
r_{1}^{n}\equiv e^{-\Gamma _{1}t/2},\ \ \ r_{2^{n}}\equiv e^{-\Gamma _{2}t/2}
\end{equation}%
where $t\equiv n\tau $ and $\Gamma _{1},\Gamma _{2}$ correspond to long and
short lifetime decay parameters respectively, then the various constants can
always be adjusted to get agreement with the standard Kaon survival
intensity functions%
\begin{eqnarray}
I\left( K^{0}\right)  &=&\tfrac{1}{4}(e^{-\Gamma _{1}t}+e^{-\Gamma
_{2}t}+2e^{-(\Gamma _{1}+\Gamma _{2})t/2}\cos \Delta mc^{2}t/\hbar ),  \notag
\\
I\left( \bar{K}^{0}\right)  &=&\tfrac{1}{4}(e^{-\Gamma _{1}t}+e^{-\Gamma
_{2}t}-2e^{-(\Gamma _{1}+\Gamma _{2})t/2}\cos \Delta mc^{2}t/\hbar )  \notag
\\
&&
\end{eqnarray}%
for pure $K^{0}$ decays. Here $\Delta m$ is proportional to the proposed
mass difference between the hypothetical $K_{1}^{0}$ and $K_{2}^{0}$ $``$%
particles$\textquotedblright $, which are each $CP$ eigenstates and are
supposed to have $CP$ conserving decay channels. From the SSQM approach,
such objects do not exist. Instead, they are simply manifestations of
different possible superpositions of $K^{0}$ and $\bar{K}^{0}$ labstates,
which are physically realizable via the strong interactions given above.

\section{Concluding remarks}

In this paper, it has been shown how signal-state quantum mechanics gives an
instrumentalist description of particle decays and the quantum Zeno effect,
consistent with Heisenberg's approach to quantum mechanics. It provides an
alternative description of quantum processes with a novel interpretation of
quantum wave-functions and avoids any reliance on the metaphysical concepts
of system under observation and continuous time. Instead of thinking about
elementary particles as strange, non-classical objects which can sometimes
appear to be waves and sometimes particles, we can choose instead to think
only of how laboratory apparatus responds to physical manipulation. This is
surely a more correct way to discuss physics, rather than in terms of
classically motivated and suspect objectification.

\

\

\begin{center}
\textbf{Acknowledgements}
\end{center}

GJ is grateful to Andrei Khrennikov for an invitation to the Foundations of
Probability and Physics-4 Conference in V\"{a}xj\"{o}, Sweden (2006) which
stimulated this line of work, and to Ariel Caticha for his interest and
helpful discussions during that meeting.

\newpage

\begin{center}
\textbf{REFERENCES}
\end{center}

\noindent Bender, C. M., Milton, K. A., Sharp, D. H., Simmons Jr., L. M.,
and Strong, R. (1985). Discrete-time quantum mechanics. \emph{Physical
Review D} \textbf{32}, 1476--1485.

\

\noindent Bjorken, J. D. and Drell, S. D. (1965). \emph{Relativistic Quantum
Fields,} McGraw-Hill Inc.

\

\noindent Bollinger, J. J., Itano, W. M., Heinzen, D. J., and Wineland, D.
J. (1990). Quantum Zeno effect. \emph{Physical Review A} \textbf{41},
22952300.

\

\noindent Caldirola, P., (1978). The chronon in the quantum theory of the
electron and the existence of heavy leptons. \emph{Il Nuovo Cimento} \textbf{%
45}, 549--579.

\

\noindent Eakins, J. and Jaroszkiewicz, G. (2005). \emph{A quantum
computational approach to the quantum universe}, in \emph{New Developments
in Quantum Cosmology Research}, \emph{Horizons in World Physics series }%
\textbf{247}, A. Reimer editor, Nova Science Publishers, Inc. New York.

\

\noindent Gell-Mann, M. and Pais, A. (1995). Behavior of neutral particles
under charge conjugation. \emph{Physical Review} \textbf{97}, 1387--1389.

\

\noindent Golden, S. (1992). Irreversible evolution of isolated quantum
systems and a discreteness of time. \emph{Physical Review A} \textbf{46},
6805--6816.

\

\noindent Heisenberg, W. (1927). \"{u}ber den anschaulichen inhalt der
quanten theoretischen kinematik und mechanik. \emph{Z. Physik} \textbf{43}%
,172--198.

\

\noindent Jaroszkiewicz, G. and Norton, K. (1997). Principles of discrete
time mechanics: I. particle systems. \emph{Journal of Physics A:
Mathematical and General} \textbf{30}, 3115--3144.

\

\noindent Jaroszkiewicz, G. and Eakins, J. (2006). \emph{Bohr-Heisenberg
reality and system-free quantum mechanics}, \emph{quant-ph/0606143}.

\

\noindent Jaroszkiewicz, G. and Ridgway-Taylor, J. (2006). Quantized
detector representation of quantum optical networks. \emph{International
Journal of Modern Physics B} \textbf{20}, 1382--1389.

\

\noindent Jordan, P. and Wigner, E. P. (1928). \"{U}ber das Paulische \"{A}%
quivalenzverbot, \emph{Zeitschrift f\"{u}er Physik} \textbf{47}, 631--651.

\

\noindent Leighton, R. B., Feynman, R. P, and Sands, M. (1966). \emph{The
Feynman Lectures on Physics: Quantum Mechanics}, volume III, Addison-Wesley
Publishing Company.

\

\noindent Meschini, D. (2006). Planck-scale physics: Facts and belief. \emph{%
Foundations of Science}.

\

\noindent Misra, B. and Sudarshan, E. C. G. (1977). The Zeno's paradox in
quantum theory. \emph{Journal of Mathematical Physics} \textbf{18}, 756--763.

\

\noindent Peres, A. (1993). \emph{Quantum Theory: Concepts and Methods,}
Kluwer Academic Publishers.

\

\noindent Wigner, E. P. (1949). Invariance in physical theory. \emph{%
Proceedings of the American Philosophical Society} \textbf{93}, 521--526.

\

\noindent Wu, L. A. and Lidar, D. A. (2002). Qubits as parafermions. \emph{%
Journal of Mathematical Physics} \textbf{43}, 4506--4525.

\

\noindent Yamamoto, H. (1984). Quantum field theory on discrete space-time.
\emph{Physical Review D} \textbf{30}, 1727--1732.

\

\noindent Zurek, W. (2002). Decoherence and the transition from quantum to
classical - revisited. \emph{Los Alamos Science} \textbf{27}, 2--24.

\end{document}